\documentclass[twocolumn,pre,amsmath,amssymb,floatfix,balancelastpage]{revtex4-2}

\usepackage{graphicx}
\usepackage{dcolumn}
\usepackage[colorlinks=true,citecolor=blue,linkcolor=blue,urlcolor=blue]{hyperref}
\usepackage{bm}
\usepackage{amsthm}
\usepackage{mathrsfs}
\usepackage{xfrac}
\usepackage{dutchcal}
\usepackage{xcolor,empheq}
\usepackage{multirow}%
\usepackage{minibox}%
\usepackage{grfext}
\usepackage{tikz}

\AtBeginDocument{\usepackage{booktabs}}

\DeclareMathAlphabet{\mathdutchcal}{U}{dutchcal}{m}{n}
\SetMathAlphabet{\mathdutchcal}{bold}{U}{dutchcal}{b}{n}
\DeclareMathAlphabet{\mathdutchbcal}{U}{dutchcal}{b}{n}

\newcommand{\multiClass}{\alpha}
\newcommand{\multi}{\mathdutchcal{m}}
\newcommand{\Multi}{\mathdutchcal{M}}

\newcommand{\ens}[1]{\left\langle #1 \right\rangle}
\newcommand{\EE}{\mathscr E}
\newcommand{\ee}{\varepsilon}

\definecolor{light_blue}{rgb}{0.8, 0.888151, 1.}
\colorlet{shadecolor}{light_blue}
\newcommand*\mybluebox[1]{ 
\colorbox{shadecolor}{\hspace{1em}#1\hspace{1em}}} 
\newcommand{\LFour}{
\begin{tikzpicture}
\draw[color=white,fill=black,radius=2pt] (0pt,0pt) circle ;
\draw[color=white,fill=black,radius=2pt] (4pt,0pt) circle ;
\draw[color=white,fill=black,radius=2pt] (8pt,0pt) circle ;
\draw[color=white,fill=black,radius=2pt] (0pt,4pt) circle ;
\end{tikzpicture}
}
\newcommand{\TFour}{
\begin{tikzpicture}
\draw[color=white,fill=black,radius=2pt] (0pt,0pt) circle ;
\draw[color=white,fill=black,radius=2pt] (4pt,0pt) circle ;
\draw[color=white,fill=black,radius=2pt] (8pt,0pt) circle ;
\draw[color=white,fill=black,radius=2pt] (4pt,4pt) circle ;
\end{tikzpicture}
}
\newcommand{\IFour}{
\begin{tikzpicture}
\draw[color=white,fill=black,radius=2pt] (0pt,0pt) circle ;
\draw[color=white,fill=black,radius=2pt] (4pt,0pt) circle ;
\draw[color=white,fill=black,radius=2pt] (8pt,0pt) circle ;
\draw[color=white,fill=black,radius=2pt] (12pt,0pt) circle ;
\end{tikzpicture}
}
\newcommand{\IThreeParentA}{
\raisebox{-0.25\height}{
\begin{tikzpicture}
\draw[color=white,fill=black,radius=2pt] (0pt,0pt) circle ;
\draw[color=white,fill=black,radius=2pt] (4pt,0pt) circle ;
\draw[color=white,fill=black,radius=2pt] (8pt,0pt) circle ;
\draw[color=white,fill=orange,radius=1pt] (0pt,4pt) circle ;
\draw[color=white,fill=orange,radius=1pt] (8pt,4pt) circle ;
\draw[color=white,fill=orange,radius=1pt] (0pt,-4pt) circle ;
\draw[color=white,fill=orange,radius=1pt] (8pt,-4pt) circle ;
\end{tikzpicture}
}}
\newcommand{\IThreeParentB}{
\raisebox{-0.25\height}{
\begin{tikzpicture}
\draw[color=white,fill=black,radius=2pt] (0pt,0pt) circle ;
\draw[color=white,fill=black,radius=2pt] (4pt,0pt) circle ;
\draw[color=white,fill=black,radius=2pt] (8pt,0pt) circle ;
\draw[color=white,fill=orange,radius=1pt] (4pt,4pt) circle ;
\draw[color=white,fill=orange,radius=1pt] (4pt,-4pt) circle ;
\end{tikzpicture}
}}
\newcommand{\IThreeParentC}{
\raisebox{-0.0\height}{
\begin{tikzpicture}
\draw[color=white,fill=black,radius=2pt] (0pt,0pt) circle ;
\draw[color=white,fill=black,radius=2pt] (4pt,0pt) circle ;
\draw[color=white,fill=black,radius=2pt] (8pt,0pt) circle ;
\draw[color=white,fill=orange,radius=1pt] (-4pt,0pt) circle ;
\draw[color=white,fill=orange,radius=1pt] (12pt,0pt) circle ;
\end{tikzpicture}
}}
\begin{document}
\preprint{}
\title{Stochastic Processes and Statistical Mechanics}
\author{Themis Matsoukas}
\email{tmatsoukas@icloud.com}
\affiliation{}
\date{\today}
\begin{abstract}

Statistical thermodynamics delivers the  probability distribution of the equilibrium state of matter through the constrained maximization of a special functional, entropy. Its elegance and enormous success have led to numerous attempts to decipher its language and make it available to problems outside physics, but a formal generalization has remained  elusive. Here we show how the formalism of thermodynamics can be applied to any stochastic process. 
We view a stochastic process as a random walk on the event space of a random variable that produces a feasible distribution of states. The set of feasible distributions obeys thermodynamics: the most probable distribution is the canonical distribution, it maximizes the functionals of statistical mechanics, and its parameters satisfy the same Legendre relationships. Thus the formalism of thermodynamics --no new functionals beyond those already encountered in statistical physics-- is shown to be a stochastic calculus, a universal language of probability distributions and stochastic processes.

\end{abstract}
\maketitle


\section{Thermodynamic Formalism}

The machinery of statistical thermodynamics can be summarized as follows \citep{Ruelle:Book-Thermodynamic-Formalism2004}: the central quantity of interest is a probability distribution and is determined through the maximization of a special functional. The result of this maximization expresses the distribution in exponential form with parameters that are related to each other via the Legendre transformation. 
The distribution in question is the probability of microstate of a fixed number of interacting particles within given volume with fixed mean energy. The functional to be maximized is
\begin{equation}
\label{statmech_functional_Gamma}
   \mathcal P[p] = 
   -\int_\Gamma p(\Gamma) \log p(\Gamma) d\Gamma
   \leq\log\omega(\bar\epsilon) ,
\end{equation}
and its maximum is the log of the microcanonical partition function. Maximization is done with respect to all probability distributions $p(\Gamma)$ with mean energy $\bar \epsilon$ and the result is the probability of microstate:
\begin{equation}
   p^*(\Gamma) = \frac{e^{-\beta\epsilon(\Gamma)}}{q} . 
\end{equation}
The parameters $\beta$ (inverse temperature), $q$ (canonical partition function) and $\omega$ (microcanonical partition function) satisfy the relationships 
\begin{gather}
\label{statmech_legendre1}
   \log\omega(\bar \epsilon) = \beta \bar \epsilon + \log q(\beta) , 
   \\
\label{statmech_legendre2}
   \bar \epsilon = -\frac{\partial\log q}{d\beta},\quad
   \beta = \frac{\partial\log\omega}{\partial\bar \epsilon} , 
\end{gather}
which express the fact that $\log\omega(\bar\epsilon)$ and $\log q(\beta)$ are Legendre pairs. 

This formalism is not limited to the probability of microstate. It may be applied to other distributions of the ensemble, the energy distribution, for example.  Given a distribution of microstates, $p(\Gamma)$,  the energy distribution is $f(\epsilon) = w(\epsilon) p(\Gamma)$, where $w(\epsilon) = d\Gamma/d\epsilon$ is the density of microstates per unit energy, or more colloquially, the number of microstates with energy $\epsilon$. Thus the equilibrium distribution of energy is
\begin{equation}
\label{statmech_mpd}
   f^*(\epsilon) = w(\epsilon)  \frac{e^{-\beta\epsilon}}{q} , 
\end{equation}
and is accompanied by the same Legendre relationships in  Eqs.\ (\ref{statmech_legendre1})--(\ref{statmech_legendre2}). 
The energy distribution maximizes its own functional, which we identify by setting $p(\Gamma) = f(\epsilon)/w(\epsilon) $ and $p(\Gamma) d\Gamma = f(\epsilon)d\epsilon$ in Eq.\ (\ref{statmech_functional_Gamma}):
\begin{equation}
\label{statmech_functional}
   \mathcal F[f] = 
   -\int f(\epsilon) \log \frac{f(\epsilon)}{w(\epsilon)} d\epsilon 
   \leq\log\omega(\bar\epsilon) . 
\end{equation}
Since this inequality turns into an exact equality only for $f=f^*$, the determination of the equilibrium energy distribution is once again reduced to a variational problem, the constrained maximization of functional $\mathcal F$. All details regarding the physical system are relegated to a single function, $w(\epsilon)$, the number of microstates with energy $\epsilon$, whose determination depends on the physics that govern the system, whether classical mechanics, quantum mechanics, or other. 
 
This formalism, concise and elegant, has proven very successful in physics in the study of problems whose enormous complexity would have suggested they are intractable. It is natural to conjecture whether the same approach could be generalized to \textit{any} problem involving an unknown probability distribution. This premise, first articulated by \citet{Jaynes:PR57}, has motivated numerous attempts to apply the thermodynamic toolbox to an ever expanding range of problems outside physics that have included statistical 
inference \citep{Jaynes:83}, 
chaotic systems \citep{Beck:book1993}, 
turbulence and vehicular flow \cite{Chowdhury:PhysRep:2000,Heinz:Book:2003}, 
social sciences \citep{Schulz:book_2003,Ayres:book2009,Contucci:FrontiersPhys:2020},  
graph theory and networks \citep{Bianconi:PRL01,Albert:RMP02,Achlioptas:S09,Matsoukas:PRE:2015},
ecology and populations \citep{Demetrius:PNAS:1997,Harte:Book:2011,Matsoukas:PRE:2014,Marquet:PNAS:2017,Matsoukas:Entropy:2020}. 
In their diversity these problems share one feature in common: the central unknown is a probability distribution. Special adaptations, however, do not generalize from one problem to the next and leave us with no systematic methodology for taking statistical thermodynamics beyond physics. 
 
Equations (\ref{statmech_legendre1})--(\ref{statmech_mpd})  themselves are a direct consequence of the maximization of the functional in Eq.\ (\ref{statmech_functional}) with respect to $f$ under the constraints 
\begin{equation}
\label{statmech_constraints}
   \int f(\epsilon) d\epsilon = 1,\quad
   \int \epsilon f(\epsilon) d\epsilon = \bar \epsilon. 
\end{equation}
The maximization of this particular functional, which in statistical mechanics is presented as a postulate, has a probabilistic interpretation that is independent of physics  \citep{Matsoukas:Entropy:2019}: it identifies the most probable distribution  $f^*$ in a space of distributions obtained by biased sampling from an exponential distribution with the same mean. Bias is implemented via a functional $W$, such that the probability of accepting a random sample with distribution $f$ is proportional to $W[f]$. Distribution $f^*$ is then shown to maximize the functional in Eq.\ (\ref{statmech_functional}) and thus to satisfy Eqs.\ (\ref{statmech_legendre1})--(\ref{statmech_mpd}), while $\log w$ is shown to be the functional derivative $\delta\log W[f^*]/\delta f^*$.

Since the central element of all stochastic processes is a probability distribution, a plausible path emerges to make formal contact between thermodynamics and stochastic processes.  A discrete stochastic process may be viewed as a random walk on the event space of a stochastic variable.  A group of $N$ walkers performing a concerted walk of $G$ steps produces a \textit{feasible distribution} of the stochastic variable at time $t_G$. Our proposition is that the most probable distribution in the feasible set is \textit{the} probability distribution of the stochastic process, and that this distribution obeys thermodynamics: it maximizes the same functional as the canonical energy, thus it is of the exponential form in Eq.\ (\ref{statmech_mpd}), and satisfies the Legendre relationships in Eqs.\ (\ref{statmech_legendre1})--(\ref{statmech_legendre2}). 
The bias functional $W$ in this case must be determined by the relative probabilities of the individual trajectories that produce the sampled distribution, and these in turn  depend on the transition probabilities of the stochastic variable in question. The precise nature of these relationships is the subject of this work. 
\vspace*{6pt}

\begin{figure}[t]
\begin{center}
\includegraphics[width=3.25in]{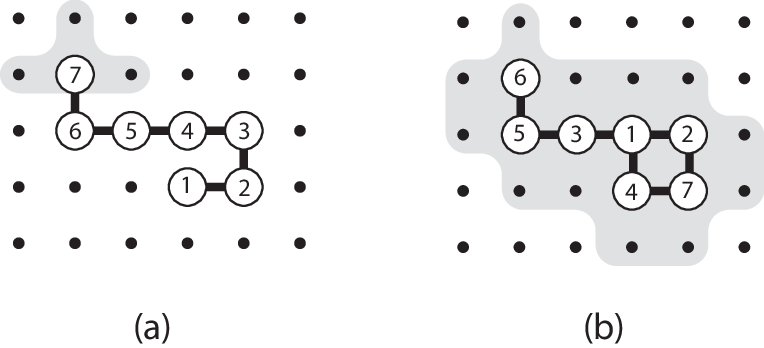}
\end{center}
\caption{Two random walks on two-dimensional lattice. (a) Walk advances from current site to a free neighbor and produces a random coil; (b) walk can advance to any free neighbor of any previously visited site and forms a random cluster. Numbers indicate the order in which a site was visited and shaded sites show the accessible states from current state. Many other walks can be constructed based on different rules.}
\label{fig_walk}
\end{figure}

\section{Stochastic Process}
\label{sct_stochastic}

\subsection{Space of Trajectories}
At its most elementary level a discrete stochastic process consists of a set $\mathscr X$ from which stochastic variable $X$ receives values, and a rule that determines the subset $\mathscr X'$ of values that are accessible from current state. In a lattice walk (Fig.\ \ref{fig_walk}a) $\mathscr X$ is the set of lattice points and $\mathscr X'$ is the set of neighboring sites of the site visited last. A more complex walk can be constructed by allowing the walker to advance to a free neighbor of any previously occupied site (Fig.\ \ref{fig_walk}b). This walk produces a lattice cluster and generates a richer set of structures that includes the random coils produced by the simpler version. 
Many more walks can be generated under various other rules.
In the development that follows will use the lattice cluster as as a concrete example for visualizing the theory with the understanding that the details of the walk are unimportant.  
 
An ordered sequence of transitions forms a \textit{trajectory}.  The generation $g$ of the trajectory is the number of transitions it contains, with $g=0$ representing the null state (no transition). The set of all trajectories that can be constructed in $g$ steps starting from the null state forms a space that we notate $\ee_g$. 
We partition this space into non overlapping subsets that we call \textit{classes} according to some property of the walk. Using the random cluster as an example, we define class as the set of clusters with the same structure, namely, clusters that can be made to coincide under translation, rotation and reflection operations on the lattice. In general, classes can be defined by any property of the trajectory, for example, radius of gyration, energy under an assumed interaction between occupied sites, and so on. 

\begin{figure}[t]
\begin{center}
\includegraphics[width=\linewidth]{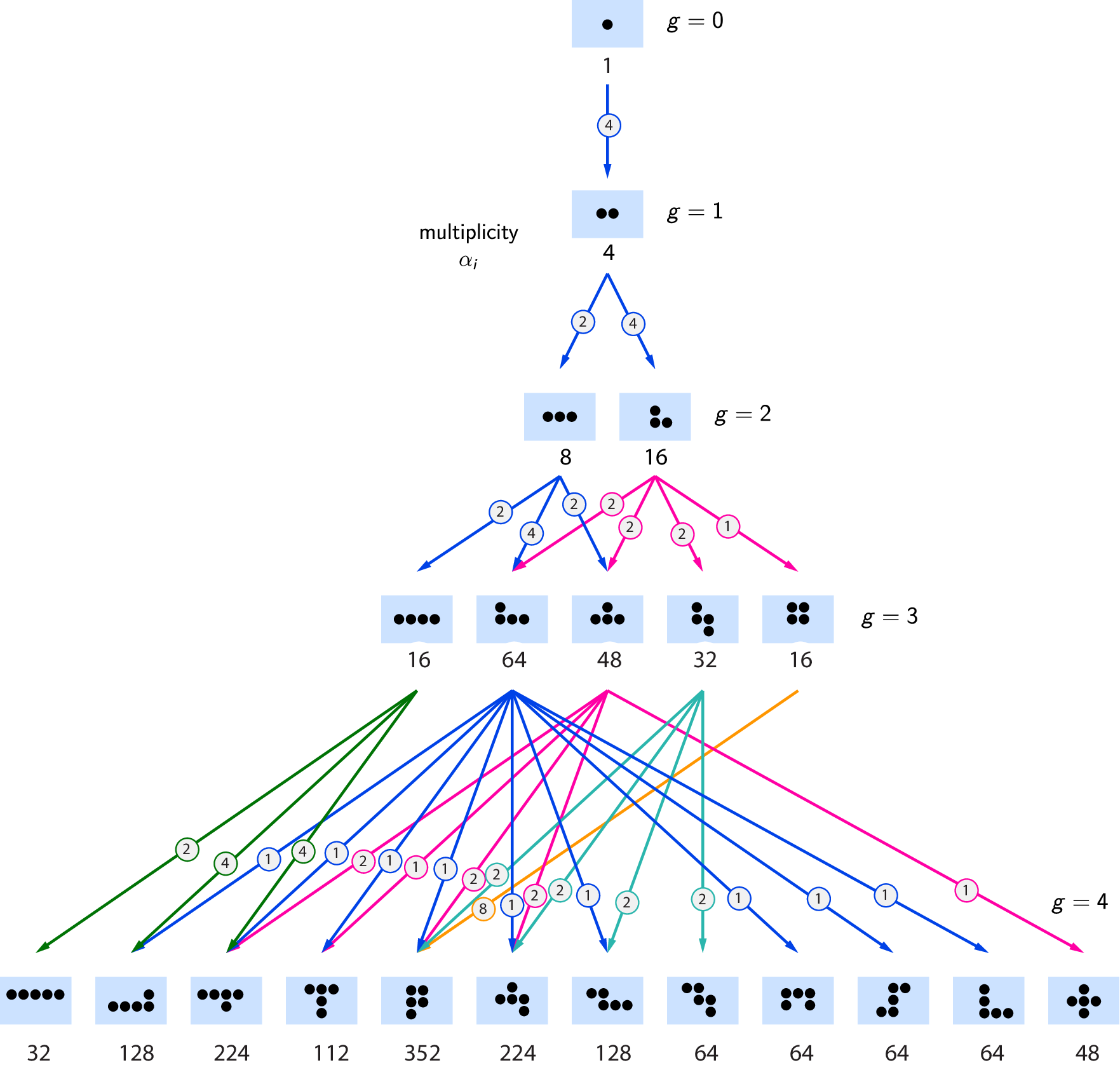}
\end{center}
\caption{Transitions of the random cluster in $2d$ square lattice up to generation $g=4$ starting with one occupied site in generation $0$ (null state). Classes are formed by clusters that are reflection or rotation images of each other. 
The propagation intensity (shown on the arrows) is equal to the number of ways the parent structure produces the offspring. For example, the linear trimer has three offspring according to the transitions $\protect\IThreeParentC\xrightarrow{2}\protect\IFour$, $\protect\IThreeParentA\xrightarrow{4}\protect\LFour$ and $\protect\IThreeParentB\xrightarrow{2}\protect\TFour$. Each parent transmits its multiplicity in proportion to the intensity of the transition. The multiplicity is written underneath each cluster and satisfies Eq.\ (\ref{multi_class}). 
}
\label{fig_rCluster_class}

\end{figure}

When a trajectory in class $i$ of generation $g-1$ undergoes a transition, it forms a new trajectory in some class $j$ of generation $g$. This amounts to a transition $i\to j$ between classes that can be represented in the form of a directed layered graph, as illustrated in Fig.\ \ref{fig_rCluster_class}. 
The graph is layered by generation and strictly directed from one generation to the next.
The parents of class $j$ in generation $g$ is the set of all classes $i$ in generation $g-1$ that emit a transition to offspring class $j$.  
We define the multiplicity $\multiClass_i$ of class as the number of trajectories it contains, and the intensity $K_{i\to j}$ of the transition, such that $\multiClass_i\, K_{i\to j}$ is the number of transitions from class $i$ to $j$. 
The multiplicity of class $j$ is the total number of transitions arriving at $j$ from all parents:
\begin{equation}
\label{multi_class}
   \multiClass_j = \sum_{i|j} \multiClass_i K_{i\to j} .
\end{equation}
The number of transitions between classes is a property of the process. All multiplicities and transition intensities in future generations can be calculated recursively  starting with $\multiClass_\text{null}=1$ in generation $g=0$. 


\vspace*{-10pt}
\subsection{Space of Distributions}
\begin{figure}[tb]
\begin{center}
\includegraphics[width=1.5in]{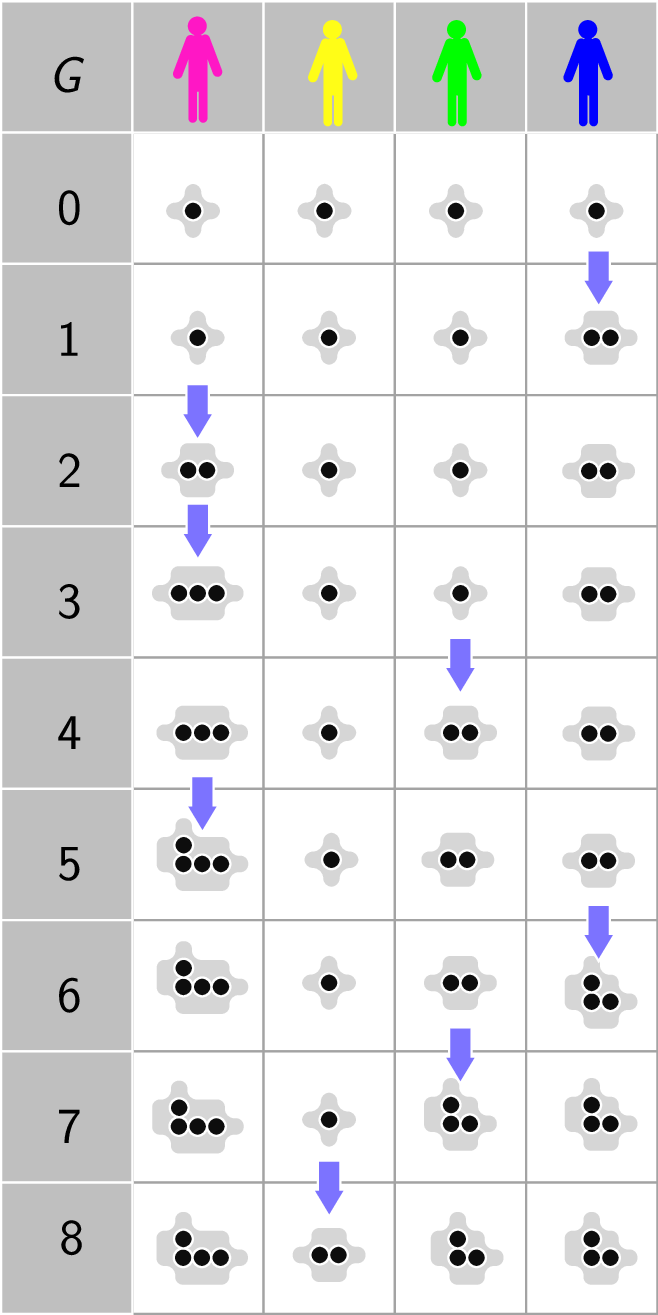}
\end{center}
\caption{Four walkers collecting samples of random clusters. At each step one walker undergoes a transition (indicated by arrows) that replaces the cluster held by the walker with one of its offspring. All transitions available to all four walkers in generation $G$ are equally probable. The sample is an ordered list of four clusters, the observable is a distribution of  classes the sample and its multiplicity is the number of ways the sample can materialize in $G$ transitions starting with all walkers in the null state (all monomers). The multiplicity of the distribution in this example after 8 steps is 1\,321\,205\,760. 
}
\label{fig_rCluster_dstr}
\end{figure}

We send  $N$ walkers on a walk on the space of classes, such that at each step one transition is implemented at random among all transitions that are possible at current state and the corresponding walker advances to a new state. Initially all walkers are at the null state (Fig.\ \ref{fig_rCluster_dstr}). 
This represents an $N$-dimensional random walk, a Markov chain whose instantaneous state is an ordered sequence of classes, configuration, that can transition to a new  configuration via a transition in a classes. 
The ordered sequence of states, starting from generation $G=0$, forms a random path.  The set of all paths that are produced in $G$ steps forms a space that we notate $\EE_{G,N}$. 
 
We form the distribution of classes in the configuration and represent it by vector $\mathbf{n}=(n_1,n_2\cdots)$, whose $i$th element is the number of walkers stationed in class $i$ (Fig.\ \ref{fig_rCluster_dstr}). All distributions in generation $G$ contain $N$ walkers and $G$ transitions and they all satisfy
\begin{equation}
\label{constraints}
   \sum n_i = N,\quad
   \sum_i g_i n_i = G,
\end{equation}
where $g_i$ is the number of transitions in class $i$, not to be confused with the multiplicity of class $\multiClass_i$. These completely define the space of feasible distributions: any distribution that satisfies the above conditions can be represented by an $N$-dimensional walk on the space of classes. 

\subsection{Multiplicity of Distribution}

We define the multiplicity $\multi(\mathbf{n})$ of distribution as the number of paths that arrive at the distribution, namely, the number of ways to assign $G$ transitions numbered in ascending order   to $N$ walkers, such that the number of walkers in class $i$ with $g_i$ transitions is $n_i$. We calculate this multiplicity by a straightforward combinatorial argument: The number of ways to assign $G$ ordered transitions to $N$ walkers such that $n_i$ walkers are in class $i$ is equal to the product $G!\, \mathbf{n!}$, where $\mathbf{n!}$ is the multinomial coefficient of vector $\mathbf{n}$. If the multiplicity of class $i$ is $\multiClass_i$, the number of ways to assign transitions increases by the factor $\multiClass^{n_1}\multiClass^{n_2}\cdots$, but this overcounts by a factor $g_i!$ per trajectory in class $i$ because only one permutation is acceptable, the one in which transitions appear in strictly ascending order. The result therefore is
\begin{empheq}[box=\mybluebox]{equation}
\label{multi_dstr}
    \multi(\mathbf{n})
    =
    G!\, \mathbf{n!}
    \prod_i\left(\frac{\multiClass_i}{g_i!}\right)^{n_i} ,
\end{empheq}
with $\mathbf{n!} = N/n_1! n_2!\cdots$. The multiplicity of distribution satisfies the propagation identity
\begin{equation}
\label{Multi_GN}
   \multi(\mathbf{n}) = 
   \sum_{\mathbf{n'}} K_{i\to j}\, n'_j\, \multi(\mathbf{n'}) ,
\end{equation}
which expresses the fact that the number of paths arriving at $\mathbf{n}$ from parent $\mathbf{n'}$ is equal to the multiplicity of the parent times the number of trajectories transmitted from that parent to $\mathbf{n}$ via transition $i\to j$. 
The sum of all multiplicities $\multi(\mathbf{n})$ is the total number of paths in $\EE_{G,N}$,
\begin{equation}
\label{Multi_propagation}
   \Multi_{G,N} 
   =
   \sum_\mathbf{n'\to n}
   K_{i\to j}\, n'_i\,  \multi(\mathbf{n'}) ,
\end{equation}
with the summation over all transitions $\mathbf{n'}\xrightarrow{i\to j}\mathbf{n}$ from generation $G-1$ to $G$. 

Equations (\ref{multi_dstr}) -- (\ref{Multi_propagation}) are structural properties of the space of classes: they arise solely from the connectivity between classes and the intensity of the transitions that connect them.

\section{Probabilities}
\subsection{Microcanonical Probability}
The set of all paths by $N$ walkers in $G$ steps forms space $\EE_{G,N}$that contains $\Multi_{G,N}$ elements. The distribution of classes represents a partitioning of this space, such that partition $\mathbf{n}$ contains $\multi(\mathbf{n})$ elements. We now assign probabilities to distributions in proportion to the number of elements they contain:
\begin{empheq}[box=\mybluebox]{equation}
\label{prob_multi}
    \Pr(\mathbf{n}|G,N) 
    = \frac{\multi(\mathbf{n})}{\Multi_{G,N}}.
\end{empheq}
This amounts to treating all paths as equiprobable; we call this probability microcanonical and note that it is properly normalized within $\EE_{G,N}$. The microcanonical probability satisfies the propagation identity
\begin{empheq}[box=\mybluebox]{multline}
\label{propagation_prob}
   \Pr(\mathbf{n}|G,N)
   =
   \frac{1}{N\ens{K}_{G-1,N}}\times
   \\ 
   \sum_\mathbf{n'} n'_i\, K_{i\to j}\, \Pr(\mathbf{n'}|G-1,N) , 
\end{empheq}
where $\ens{K}_{G-1,N}$ is the mean transition intensity from generation $G-1$ to $G$, defined as
\begin{equation}
   \ens{K}_{G-1,N}
   =
   \frac{1}{N}
   \sum_\mathbf{n'\to n} 
   n'_i K_{i\to j}\, \Pr(\mathbf{n'}),
\end{equation}
with the summation taken over all transitions between the two generations. These follow directly from the propagation of multiplicity in Eqs.\ (\ref{Multi_GN}) and (\ref{Multi_propagation}). 

Given a set of rules that define the allowable transitions between classes, all multiplicities and all transition intensities are fixed. Then the probability defined by Eq.\ (\ref{prob_multi}) satisfies Eq.\ (\ref{propagation_prob}) as an identity.  
Alternatively we may consider Eq.\ (\ref{propagation_prob}) to define $P(\mathbf{n})$ recursively from initial state $\mathbf{n}_0$ with $P(\mathbf{n}_0)=1$; in this case its solution is Eq.\ (\ref{prob_multi}). 
In this view the intensities $K_{i\to j}$ may be regarded as arbitrary functions, not necessarily tied to actual number of transitions between classes. 
%

\subsection{Partition Function}
\label{sct_thermo}
Using Eq.\ (\ref{multi_dstr}) for the multiplicity of distribution the microcanonical probability in Eq.\ (\ref{prob_multi}) becomes
\begin{empheq}[box=\mybluebox]{equation}
\label{prob_Omega_W}
    \Pr(\mathbf{n}) 
    = \frac{N!}{\Omega_{G,N}}
      \prod_i \frac{(\multiClass_i/g_i!)^{n_i}}{n_i!} , 
\end{empheq}
where 
$\Omega_{G,N}$ is the partition function that normalizes this probability,
\begin{equation}
\label{Omega_def}
   \Omega_{G,N} = 
   N!\sum_\mathbf{n} \prod_i \frac{(\multiClass_i/g_i!)^{n_i}}{n_i!} . 
\end{equation}
Notice that $\Omega_{G,N}=\Multi_{G,N}/G!$, as $G!$ cancels from both the numerator and denominator in Eq.\ (\ref{prob_multi}).
The form of Eq.\ (\ref{Omega_def}) is known as Gibbs distribution \cite{Berestycki:07} and is commonly encountered in stochastic processes and statistical mechanics \cite{Durrett:JTP99,Evans_2005,Matsoukas:PRE:2014,Berestycki:07}. As a special result we obtain the mean number of class $i$ in the ensemble of generation $G$ in closed form in terms of the partition function and the apparent multiplicity \cite{Evans_2005,Matsoukas:PRE:2014}:
\begin{equation}
\label{mean_dstr}
   \frac{\ens{n_i}}{N} 
   = 
   \frac{\multiClass_i}{g_i!}
   \frac{\Omega_{G-g_i,N-1}}{\Omega_{G,N}} . 
\end{equation}
The result is valid for all $G$ and $N$. 

Up to this point we have assigned probabilities to distributions but not to the classes themselves. Viewing the frequency $n_i/N$ of class $i$ within a sample as an estimate of the probability of class, we define the probability of class as the mean value of this ratio over the ensemble: 
\begin{gather}
   \Pr(i|G, N) = 
   \frac{\multiClass_i}{g_i!}
   \frac{\Omega_{G-g_i,N-1}}{\Omega_{G,N}}.
\end{gather}
Within the subset of classes with the same number of transitions $g_i$ the probability of class is proportional to the class multiplicity $\multiClass_i$ and this is true for any $G$ or $N$. Even as the walkers accumulate samples, the probability of class within the subset $g_i=\text{const.}$ remains constant at all times and equal to the probability of class in its own generation, as in the graph in Fig.\ \ref{fig_rCluster_class}. As a corollary we recognize Fig.\ \ref{fig_rCluster_class} as the walk of a single walker. 

\subsection*{Propagation Equations}

With Eqs.\ (\ref{Omega_def})--(\ref{prob_Omega_W}) we have recast the microcanonical probability in terms of the partition function and the sampling multiplicity. We complete the formulation by obtaining their propagation equations from one generation to the next. 
The propagation equation for the sampling multiplicity follows from Eq.\ (\ref{multi_class}) with the substitution $w_i=\multiClass_i/g_i!$:
\begin{equation} 
\label{w_propagation}
    w_i = \frac{1}{g_i}\sum_{i|j} K_{i\to j} w_j . 
\end{equation}
The propagation of the partition function is obtained from Eq.\ (\ref{Multi_propagation}) for the total multiplicity; this is solved recursively, starting from generation $G=0$ with $\Omega_{0,N}=1$, to obtain the final result in the form
\begin{empheq}[box=\mybluebox]{equation}
\label{Omega_propagation}
    \Omega_{G,N} 
    = \left(\frac{N^G}{G!}\right) \prod_{\gamma=0}^{G-1} \ens{K}_{\gamma,N} . 
\end{empheq}
To complete this discussion we present the propagation of the mean number of classes from one generation to the next. The result, whose details can be found in the Appendix, is
\begin{multline}
\label{pbe}
   \ens{n_i}_G - \ens{n_i}_{G-1}
   = 
   \\ 
   \frac{\ens{\sum_k n_k K_{k\to i} - \sum_k n_i K_{i\to k}}_{G-1}}{N\ens{K}_{G-1}}
   +
   \\ 
   \left\{
      \frac{\ens{n_i K(\mathbf{n})}_{G-1}}{\ens{K}_{G-1}}  
      - \ens{n_i}_{G-1} 
   \right\} , 
\end{multline}
%
where $K(\mathbf{n})$ is the total propagation rate in distribution $\mathbf{n}$:
\begin{gather}
   K(\mathbf{n}) = 
   \frac{1}{N}\sum_i \sum_j n_i K_{i\to j} ,  
\end{gather}
with the summation is over all transitions originating from distribution $\mathbf{n}$. 
The left-hand side of Eq.\ (\ref{pbe}) is the change in the mean number of class $i$ from generation $G-1$ to $G$. The first term on the right-hand side in the generation and depletion of class $i$ as classes transition in and out of class $i$. The last term accounts for the fact that different distributions produce offspring with different rates. In the special case that all distributions have the same number of offspring we obtain $K(\mathbf{n}) = \ens{K}$ and this term drops out. 

The results of this section, Eqs.\ (\ref{prob_Omega_W}) through (\ref{pbe}), are exact for all finite $G$ and $N$. Next we obtain asymptotic results for large $N$.

\subsection{Most Probable Distribution}

Setting $f_i=n_i/N$ and using the Stirling approximation for the factorial, the asymptotic form of the microcanonical probability in Eq.\ (\ref{prob_Omega_W}) is
\begin{equation}
\label{prob_AL}
   \Pr(f) = \Pr(\mathbf{n}) \sim e^{-N\varrho(f) + O\log 1/N} ,
\end{equation}
where $\varrho(f)$ is the microcanonical functional
\begin{empheq}[box=\mybluebox]{equation}
\label{mC_functional}
   \varrho(f) = -\sum_i f_i \log \frac{f_i}{w_i} - \log\omega ,
\end{empheq}
with $w_i = \multiClass_i/g_i!$ and $\omega = \Omega^{1/N}$. The probability in Eq.\ (\ref{prob_AL}) satisfies the large deviations principle \citep{Touchette:PR09} and in the asymptotic limit peaks sharply about its most probable element $f^*$. In this limit the most probable distribution is overwhelmingly more probable than all others. 
We obtain this distribution by maximizing the microcanonical functional with respect to all $f$ that satisfy
\begin{equation}
   \sum_i f_i = 1,\quad
   \sum_i g_i f_i = G/N \equiv \bar g . 
\end{equation}
This is the same variational problem as in familiar thermodynamics and produces the same results: the most probable distribution is the canonical distribution,
\begin{empheq}[box=\mybluebox]{equation}
\label{mpd}
    f^*_i 
    = w_i \frac{e^{-\beta g_i}}{q}
    = \frac{\multiClass_i}{g_i!} \frac{e^{-\beta g_i}}{q}, 
\end{empheq}
and its parameters $\beta$ and $q$ satisfy the Legendre relationships:
\begin{empheq}[box=\mybluebox]{gather}
\label{omega_beta_q}
    \log\omega = \beta\bar g + \log q ; \\
\label{legendre}
    \beta = \frac{\partial\log\omega}{d\bar g};\quad
    \bar g = - \frac{d\log q}{d\beta} . 
\end{empheq}
We have made full contact with statistical thermodynamics.

\section{Distinguishability and Gibbs's Paradox}
Equation (\ref{prob_Omega_W}) gives the probability of a sample collected by $N$ walkers following a sequence of transitions between classes from null initial state to current state. The same equation has an independent combinatorial interpretation: 
it is the probability to obtain  distribution $\mathbf{n}$ by random sampling from a pool of classes in which class $i$ appears $w_i$ times. The apparent multiplicity of class in this pool is $\multiClass_i/g_i!$, not $\multiClass_i$, as if transitions were indistinguishable. Yet transitions are distinguishable: they are numbered consecutively and the order in which they form a trajectory is distinct from other permutations. 
To interpret the the factorial term properly we begin with the observation that when $\multiClass_i = g_i!$, the effective multiplicity is $w_i=1$. In this special case the microcanonical functional is
\begin{equation}
   \rho(f) = -\sum_i f_i \log p_i - \log\omega . 
\end{equation}
This is the Shannon functional of distribution $f$ minus a constant. Accordingly, the most probable distribution is the maximum entropy distribution, a distribution in which the probability of class is exponential in the number of transitions contained in the class (size of cluster). 
In light of this observation the factorial may be viewed as a prior that normalizes multiplicity. 
In the absence of any specific knowledge about the process the uninformed prior is that all $g_i!$ permutations are equally probable. If we express the relationship between $\multiClass_i$ and $w_i$ as a conditional statement, 
\begin{gather}
   \multiClass_i = w_{i|g_i} g_i!, 
\end{gather}
the ratio $w_i=\multiClass_i/g_i!$ may be viewed to normalize multiplicity by the multiplicity of the uninformed prior. 

We encounter the same situation with the multiplicity of distribution. Here the uniformed prior is $G!$, the number of permutations in the order in which  walkers collect $G$ samples. The partition function $\Omega_{N,G}$ is the total multiplicity divided by the uniformed prior, which amounts to treating $\multi(\mathbf{n})/G!$ as the effective multiplicity of distribution. Division by a common factor has no effect on probabilities, which may be obtained either in terms of the actual or effective multiplicity. However, to obtain thermodynamic behavior we must work with the effective multiplicity, on which the partition function is based. This is because $\log\Omega_{G,N}$ is asymptotically homogeneous in $G$ and $N$, while $\Multi_{G,N}$ is not. 

The situation is reminiscent of the Gibbs paradox: without division by the factorial term the partition function is not homogeneous. The ad hoc division, whose sole purpose is to restore homogeneity, is justified after-the-fact on grounds of indistinguishability, and has remained a point of contention \cite{,,Jaynes:MaxEntBayesian:1992,vanLith:Entropy:2018}. 
There is no ad hoc division in our approach, indeed no paradox in the first place. The sampling multiplicity of class, $\multiClass_i/g_i!$, appears unforced in the combinatorial calculation of the multiplicity of distribution. The sampling multiplicity of distribution $\multi(\mathbf{n})/G!$ similarly appears  when we take the ratio $\multi(\mathbf{n})/\Multi_{G,N}$ in Eq.\ (\ref{prob_Omega_W}) in which the factor $G!$ drops out. The partition function is identified as the normalizing constant $\Omega_{G,N}=\Multi_{G,N}/G!$ in the denominator of the microcanonical probability and is homogeneous, as it should. 
It has, however, a similar interpretation as the ratio $\multiClass_i/g_i!$: knowing only that $G$ distinguishable transitions must be assigned, the uninformed prior is that all $G!$ permutations are possible. However it is only the permutation in ascending order that represents a valid trajectory.


\begin{figure}
\begin{center}
\includegraphics[width=3.25in]{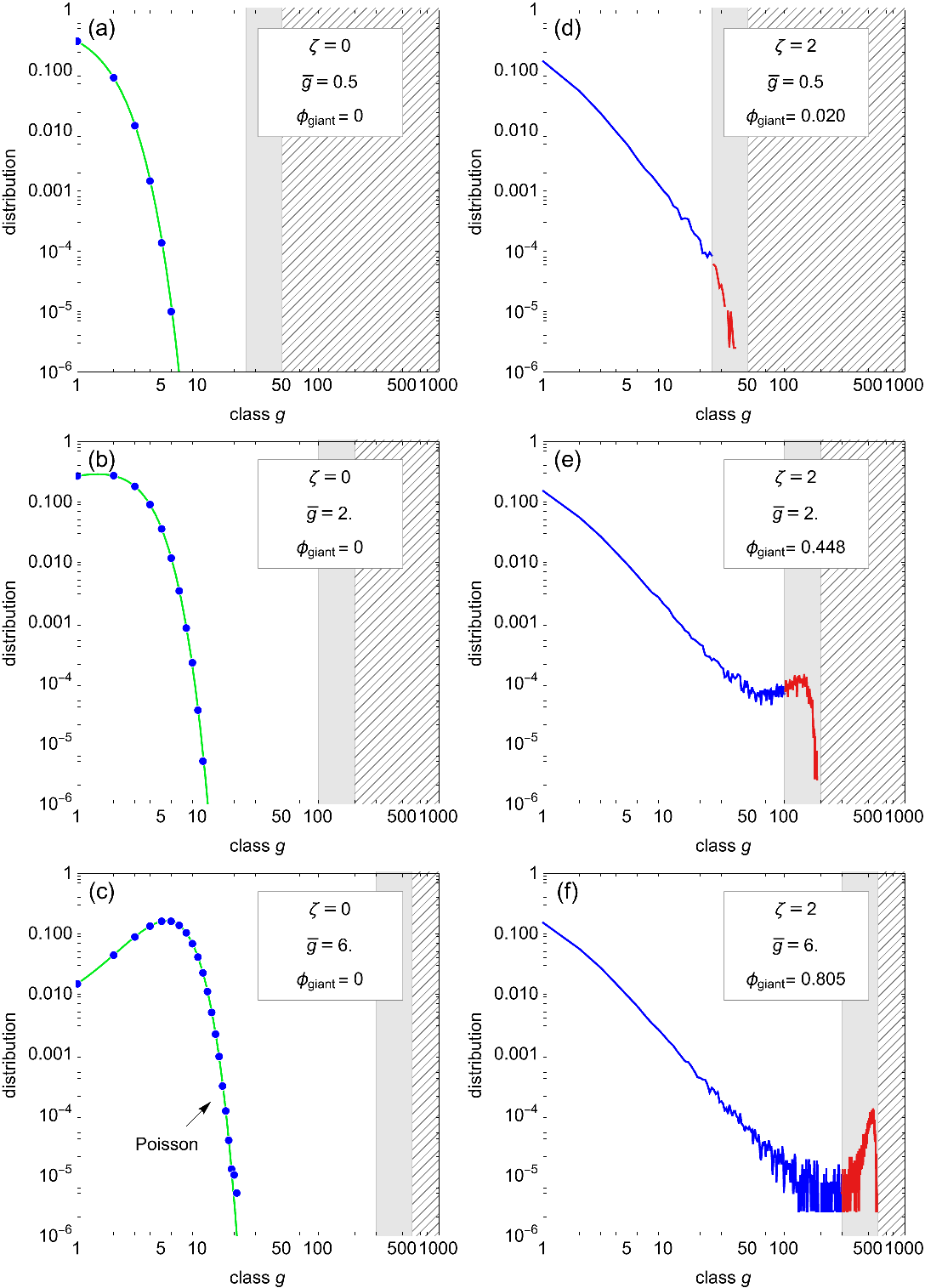}
\end{center}
\caption{Monte Carlo snapshots of 100 walkers with  power-law transition intensity, $K_g=g^\zeta$. (a)--(c) $\zeta=0$: In this case $\beta=\log/\bar g$, $q=e^{-\bar g}$, $w_g = 1/g!$, and the most probable distribution is Poisson with rate $\bar g$ (solid line); it is a stable distribution fully contained in the region $0\leq g \leq G/2$ (points are from MC simulation). (e)--(f): The population forms a giant component in the shaded region $G/2<g\leq G$ (the hatched region $g>G$ is inaccessible). The system consists of two phases, a dispersed population of walkers, and a giant component. As the process advances the giant component accumulates an increasing fraction of transitions, whereas the dispersed population remains nearly unchanged. (All lines are from MC simulations.) }
\label{fig_dstr}
\end{figure}

\section{Stability and Phase Transitions}
\label{sct_stability}
The constrained maximization that leads to the canonical form of the most probable distribution implies that the microcanonical partition function is concave in $\bar g = G/N$:
\begin{gather}
\label{stability}
   \frac{\partial^2\omega}{\partial\bar g^2} \leq 0 .  
\end{gather}
When this condition is violated, a thermodynamic system undergoes phase splitting. We demonstrate this with an example. 

For the purposes of this demonstration we define class as the set of all trajectories with the same number of transitions. In the cluster analogy, class $g$ is all clusters that contain $g+1$ monomers, starting with one monomer in the null state. We call this all encompassing partition unitary class. 
Transitions between classes are represented by
\begin{equation*}
   (g-1) \xrightarrow{K_{g-1\to g}} (g)
\end{equation*}
whose intensity is a function of $g$. Despite its simplicity this model relates to some important problems in statistical physics.  This is the so-called one-step  \citep{vanKampen} or zero-range process \citep{Evans_2005}, albeit constrained to advance in the forward direction; if we interpret $g$ as number of ``particles,'' the model represents cluster growth by monomer addition, a stochastic process with relevance to polymer growth, colloidal systems and crystallization \citep{Overbeek:AdvCollIntSci1982,Matsoukas:PRE06}. This process has no stationary solution. 

For analytic convenience we take the number of transitions to be of the power-law form $K_{g-1\to g} = g^\zeta$. In this case the partition function, its parameters $\beta$ and $q$, and the multiplicity of class are readily obtained:
\begin{gather}
   \Omega_{G,N} = 
      \left(\frac{N^G}{G!}\right)^{\zeta-1}
      \binom{G+N-1}{N-1}^\zeta;\\
   \beta = \log\frac{(\bar g+1)^\zeta}{\bar g}, \quad
   q= (\bar g+1)^\zeta\, e^{(\zeta-1)\bar g}, 
   \\
   w_g = (g!)^{\zeta-1}. 
\end{gather}
These are exact for $\zeta=0$ and $\zeta = 1$ and may serve as an approximation for arbitrary $\zeta$. 
With $\zeta=0$ the mean distribution from Eq.\ (\ref{mean_dstr}) is Poisson with rate $\bar g = G/N$. The same result can be obtained independently by straightforward combinatorics (at all times all walkers are equally probable to receive a transition), which serves nicely as a check of the thermodynamic treatment used here. For $\zeta=1$ the solution is exponential but we move on as we are interested to explore instability and phase transitions. 

The stability condition in Eq.\ (\ref{stability}) now becomes
\begin{gather}
   \bar g (\zeta-1)-1 \leq 0 . 
\end{gather}
For $\zeta<1$ it is satisfied for all $\bar g$: the system is stable and its most probable distribution is given by Eq.\ (\ref{mpd}). 
For $\zeta>1$ it is satisfied only in the region $\bar g < 1/(\zeta-1) \equiv g^*$ and is violated when $\bar g$ exceeds the critical value $g^*$. 
The nature of the instability is best illustrated via Monte Carlo simulation by comparing the cases $\zeta=0$ and $\zeta=2$  (Fig.\ \ref{fig_dstr}). With $\zeta=0$ the most probable distribution is Poisson with rate $\bar g = G/N$ and is fully contained in the region $0<g<G/2$.
For $\zeta=2$ the system is unstable once $\bar g>1$. Past the stability limit the population develops a new mode in the region $G/2<g\leq G$ that represents a giant component, a single walker that absorbs a finite fraction $\phi$ of all transitions \cite{Matsoukas:SciRep:2015,Matsoukas:Entropy:2020}. This system consists of two phases: a single giant component that accumulates an increasing fraction of the total number of transitions, and a dispersed population whose distribution is contained in the region $0\leq g \leq G/2$ and whose fraction of transitions it contains diminishes with time. Thus the state begins as a single dispersed phase and transitions to a two-phase system through the accumulation of transitions by the giant component. 

The giant component is a subject of active interest in graph theory \citep{Bollobas:Book:2001,Strogatz:N01,Newman:PRL09,Spencer:NAMS10,Buldyrev:N10}. Here we have encountered it in its simplest manifestation (number of transitions/nodes in the component) and have shown its emergence to be a phase transition in the standard thermodynamic sense.

\section{Closing Remarks}

We have formulated a theory of stochastic processes in which \textit{probability} is defined in terms of \textit{multiplicity}, an extensive property of the stochastic processes, itself well-defined, that  satisfies a forward propagation equation such that all multiplicities may be calculated systematically from the initial state. At the heart of the theory is an $N$-dimensional walk that samples a phase space of distributions, each distribution representing a possible sample of classes after $G$ steps. The set of distributions with the same number of transitions forms an ensemble $\mathscr E_{G,N}$ that we call \textit{microcanonical}, in the recognition of the fact that all \textit{configurations} in generation $G$, namely, all ordered samples of $N$ classes with a combined number of $G$ transitions, are equally probable. This equal a priori condition defines all probabilities on $\mathscr E_{G,N}$. Thermodynamics arises in the asymptotic limit $N\to\infty $ and in this limit we obtain full contact with statistical thermodynamics. 

\subsection*{What is Entropy?}

The central point of contact between this treatment and statistical mechanics is the thermodynamic functional in Eq.\ (\ref{mC_functional}), whose maximization produces the canonical form of the probability distribution. This functional can be written as
\begin{equation}
   \rho(f) 
   = -\sum_i f_i\ln f_i 
     +\sum_i f_i\ln w_i
     - \ln\omega . 
\end{equation}
In this form it consists of three terms all of which have been associated with ``entropy'' in various contexts: the first term is the Shannon entropy functional of distribution $f_i$; the second term is the mean of the log of multiplicity, or the mean of what we may call Boltzmann entropy of class $i$; and the last term is the log of the microcanonical partition function, which in statistical mechanics is identified with the thermodynamic entropy of the ensemble. Inserting the most probable distribution in the microcanonical functional yields the relationship among these functionals:
\begin{empheq}[box=\mybluebox]{gather}
   -\sum_i f^*_i\ln f^*_i 
   +\sum_i f^*_i\ln w_i
   = \ln\omega .
   \vphantom{\sum^i}
\end{empheq}
Here the first term is the Gibbs entropy of the ensemble. With $w_i=1$ we obtain the equivalence between Gibbs entropy and the log of the microcanonical partition function, or between the Boltzmann entropy and the microcanonical partition function if $f_i^*$ is a Kronecker delta--but these are special results.  It is tempting to read this equation as ``Gibbs entropy plus Boltzmann entropy equals the entropy of the ensemble'' but we prefer to stay away from the overused term ``entropy'' and adopt a neutral terminology instead:  ``Boltzmann-Gibbs-Shannon  (BGS) functional'' for the first term on the left-hand side, ``mean log of multiplicity'' for the second term, and ``log of the microcanonical partition function'' for the term on the right. 
Together, the BGS functional and the mean log of multiplicity produce the log of the microcanonical function, from which all properties of the ensemble are obtained in the asymptotic limit. 

\section{Conclusions}
\label{sct_summary} 
We now recognize the formalism of thermodynamics as a general language for stochastic processes, a language that makes no reference to \textit{stationarity}, \textit{equilibrium} or \textit{reversibility}, free of postulates and independent of physical theories, a probabilistic calculus that stands above controversies that arise from competing interpretations of physical phenomena. 
The ensemble is the set of feasible distributions and the probability measure is their multiplicity, as measured by the number of paths that lead to the distribution. The variational principle that gives rise to thermodynamics is the condition that identifies the most probable distribution in the feasible set. It is the distribution that in the asymptotic limit becomes \textit{the} probability distribution of the stochastic process. 
The ensemble probability is expressed in terms of the sampling multiplicity $w_i$, which in turn is determined by the transition intensities $K_{i\to j}$. 
If multiplicity is known, or as in the case of statistical mechanics, postulated, we  obtain the most probable distribution directly. Alternatively, multiplicities must first be computed from the transition intensities as in the example of Section \ref{sct_stability}. 
Thus we have the convergence between the two fundamental approaches in statistical mechanics: the kinetic approach, pioneered by \citet{Boltzmann:GasTheory1995}, and the ensemble approach,  formulated by \citet{Gibbs:Book1981}. 

It remains an open question as to why this calculus performs so spectacularly when applied to real physical systems composed of enormous numbers of constituent parts and governed by very complex interactions. By separating the mathematical part from the underlying physics the hope is that we may come closer to the answer. 

\bibliography{tm,statmech,particles} 
\bibliographystyle{apsrev}

\appendix
\section{Derivation of Eq.\ (\ref{pbe})}
The mean number of times class $k$ appears in the ensemble of all distributions with $G$ transitions is
\begin{equation}
\label{app_mean_nk_def}
   \ens{n_k}_G 
   = \sum_\mathbf{n} n_k \Pr(\mathbf{n}) ,
\end{equation}
where $n_k$ is the number of of times class $k$ appears in distribution $\mathbf{n}$ of the ensemble, $\Pr(\mathbf{n})$ is the probability of distribution and the summation is over all distributions in the ensemble. 
To obtain the propagation equation for $\ens{n_i}_G$ we use the propagation equation for $\Pr(\mathbf{n})$ in Eq.\ (\ref{propagation_prob}), which we write in the condensed form
\begin{equation}
\label{app_Prob_Prop}
   \Pr(\mathbf{n})
   =
   \frac{1}{N\ens{K}_{G-1,N}}
   \sum_\mathbf{n'\to n} n'_i\, K_{i\to j}\, \Pr(\mathbf{n'}) ,    
\end{equation}
with the understanding that $\mathbf{n}$ belongs in the ensemble $(G,N)$ and $\mathbf{n'}$ is a parent distribution in the ensemble $(G-1,N)$. This summation is over all parents $\mathbf{n'}$ of $\mathbf{n}$. 
Combining with Eq.\ (\ref{app_mean_nk_def}) we get
\begin{equation}
   \ens{n_k}_G 
   = 
   \frac{1}{N\ens{K}_{G-1s}}
   \sum_\mathbf{n}\sum_\mathbf{n'}
   n_k n'_i K_{i\to j} \Pr(\mathbf{n'}) . 
\end{equation}
Now the summation is over all transitions $\mathbf{n'}|G-1,N \to \mathbf{n}|G,N$weighted by the probability of $\mathbf{n'}$, it is therefore an ensemble average over the parent ensemble. Since $K_{i\to j}$ is non zero only for transitions that produce class $k$, the above result can be expressed in the equivalent form
\begin{multline}
\label{app_pbe_0}
   \ens{n_k}_G 
   =    
   \frac{1}{N\ens{K}_{G-1,N}}
   \times
   \\
   \ens{
      \sum_i\sum_j (n'_k - \delta_{k,i}+\delta _{k,j}) n'_i K_{i\to j} , 
   }_{G-1}
\end{multline}
where $\delta_{i,j}$ is Kronecker's delta. 
Calculating the the double summations individually over each term we have:
\begin{gather}
   \ens{\sum_i\sum_j n'_k n'_j K_{i\to j}}_{G-1}
   = 
   \ens{n'_k \sum_i\sum_j n'_j K_{i\to j}}_{G-1}
\end{gather}
\begin{gather}
   \ens{\sum_i\sum_j\delta_{k,i}n'_i K_{i\to j}}_{G-1}
   =
   \ens{\sum_j n'_k K_{k\to j}}_{G-1}
\end{gather}
and
\begin{gather}
   \ens{\sum_i\sum_j\delta_{k,j}n'_i K_{i\to j}}_{G-1}
   =   
   \ens{\sum_i n'_i K_{i\to k}}_{G-1}
\end{gather}
Returning these expressions into Eq.\ (\ref{app_pbe_0}) the result is
\begin{multline}
   \ens{n_k}_G
   = 
   \frac{\ens{n'_k K(\mathbf{n'})}_{G-1}}{\ens{K}_{G-1}}
   +
   \\ 
   \frac{
   \ens{
   \sum_i n'_i K_{i\to k}
   -
   \sum_i n'_k K_{k\to i}
   }_{G-1}
   }
   {N \ens{K}_{G-1}}
   ,
\end{multline}
where $K(\mathbf{n'})$ is the propagation rate of parent $\mathbf{n'}$ per walker:
\begin{gather}
   K(\mathbf{n'}) = \frac{\ens{\sum_i\sum_j n'_j K_{i\to j}}}{N}
\end{gather}
The last step is to subtract $\ens{n_k}_{G-1}$ from both sides of the equation:
\begin{multline}
\label{app_pbe}
   \ens{n_k}_G - \ens{n_k}_{G-1}
   = 
   \\ 
   \frac{\ens{\sum_i n_i K_{i\to k} - \sum_i n_k K_{k\to i}}_{G-1}}{N\ens{K}_{G-1}}
   +
   \\ 
   \left\{
      \frac{\ens{n_k K(\mathbf{n'})}_{G-1}}{\ens{K}_{G-1}}  
      - \ens{n_k}_{G-1} 
   \right\} . 
\end{multline}
This is Eq.\ (\ref{pbe}) in the text.

As noted in the text, the first term on the left-hand side is the usual population balance equation on the generation and depletion of class $k$, except that is is expressed as an ensemble average over all distributions in the parent ensemble. The second term arises from the fact that the propagation rate $K(\mathbf{n'})$ is not necessarily the same for all parents. If it is, then $K(\mathbf{n'}) = \ens{K}_{G-1}$ and the second term drops out.

\subsection*{Asymptotic Limit -- No Phase Transition}

In the asymptotic limit and in the absence of phase transitions both the parent and offspring ensembles reduce to their respective most probable distributions. In practice we may drop the ensemble averages and make the replacements
\begin{gather}
   \mathbf{n} \to \mathbf{n}^*_{G},\quad
   \mathbf{n'} \to \mathbf{n}^*_{G-1} \\ 
   K(\mathbf{n}')\to K(\mathbf{n}^*_{G-1})\to
   \ens{K}_{G-1} = K^*_{G-1}  . 
\end{gather}
Under these conditions the second term in Eq.\ (\ref{app_pbe}) drops out regardless of whether $K(\mathbf{n'})$ is strictly the same for all parent distributions or not and the propagation equation becomes
\begin{multline}
\label{app_pbe_1}
   n^*_{k|G} - n^*_{k|G-1}
   =
   \\ 
   \left. 
   \frac 
   { \sum_i n^*_i K_{i\to k} - \sum_i  n^*_k K_{k\to i} }
   {N K^*}
   \right|_{G-1} . 
\end{multline}
The product $N K^*$ is the total propagation from ensemble $(G-1,N)$ to $(G,N)$, namely, the number of trajectories transmitting form the parent ensemble to offspring. If we interpret this to mean ``rate,''  as in ``transmissions per unit time,'' its inverse is the mean time to transition between ensembles:
\begin{gather}
   \frac{1}{N K^*}\Big|_{G-1} = \Delta t\Big|_{G-1\to G} . 
\end{gather}
With this final substitution Eq.\ (\ref{app_pbe_1}) becomes
\begin{gather}
    \left.\frac{\Delta n^*_k}{\Delta t}\right|_{G\to G-1}
    = 
    \sum_i n^*_i K_{i\to k} - \sum_i n^*_k K_{i\to i} . 
\end{gather}
We recognize the result as the mean-field balance equation for a population undergoing transitions $i\to j$ with rate constant $K_{i\to j}$. 
It is important to note that the validity of this equation is contingent on the absence of a phase transition because only then the ensemble converges onto its most probable distribution. 
\end{document}